\documentclass[twocolumn]{aastex61}
\usepackage{amsmath,amstext}
\usepackage{natbib}
\usepackage{bm}
\usepackage{color}

\accepted{May 17, 2019}
\submitjournal{ApJL}

\shorttitle{Protostellar flares}
\shortauthors{Takasao et al.}

\begin{document}

\title{Giant protostellar flares: accretion-driven accumulation and reconnection-driven ejection of magnetic flux in protostars}

\correspondingauthor{Shinsuke TAKASAO}
\email{shinsuke.takasao@nao.ac.jp}

\author{Shinsuke TAKASAO}
\affiliation{Division of Science, National Astronomical Observatory of Japan, Mitaka, Tokyo, 181-8588, Japan}

\author{Kengo TOMIDA}
\affiliation{Department of Earth and Space Science, Osaka University, Toyonaka, Osaka, 560-0043, Japan}
\affiliation{Department of Astrophysical Sciences, Princeton University, Princeton, NJ 08540, USA}

\author{Kazunari Iwasaki}
\affiliation{Division of Science, National Astronomical Observatory of Japan, Mitaka, Tokyo, 181-8588, Japan}

\author{Takeru K. SUZUKI}
\affiliation{School of Arts \& Sciences, University of Tokyo, 3-8-1, Komaba, Meguro, Tokyo, 153-8902, Japan}



\begin{abstract}
Protostellar flares are rapid magnetic energy release events associated with formation of hot plasma in protostars. In the previous models of protostellar flares, the interaction between a protostellar magnetosphere with the surrounding disk plays crucial roles in building-up and releasing the magnetic energy. 
However, it remains unclear if protostars indeed have magnetospheres because vigorous disk accretion and strong disk magnetic fields in the protostellar phase may destroy the magnetosphere.
Considering this possibility, we investigate the energy accumulation and release processes in the absence of a magnetosphere using a three-dimensional magnetohydrodynamic simulation. Our simulation reveals that protostellar flares are repeatedly produced even in such a case. 
Unlike in the magnetospheric models, the protostar accumulates magnetic energy by acquiring large-scale magnetic fields from the disk by accretion. Protostellar flares occur when a portion of the large-scale magnetic fields are removed from the protostar as a result of magnetic reconnection. Protostellar flares in the simulation are consistent with observations; the released magnetic energy (up to $\sim 3\times 10^{38}$~erg) is large enough to drive observed flares, and the flares produce hot ejecta. The expelled magnetic fields enhance accretion, and the energy build-up and release processes are repeated as a result.
The magnetic flux removal via reconnection leads to redistribution of magnetic fields in the inner disk. 
We therefore consider that protostellar flares will play an important role in the evolution of the disk magnetic fields in the vicinity of protostars.
\end{abstract}




\section{Introduction}
Protostars are formed as a result of the gravitational collapse of cold ($\sim10$~K) molecular cloud cores \citep{mckee2007}. As the contracting gas should have finite angular momentum, circumstellar disks form as natural byproducts. Since the protostellar evolution is largely controlled by accretion through the disks \citep{matt2005,tomida2017}, it is important to reveal the star-disk interaction to understand the protostellar growth history.

Since the star formation takes place in a cold environment, it had been considered that deeply embedded protostars such as Class-0 and -I objects are irrelevant to high-energy processes. However, X-ray observations revealed that Class-I protostars emit hard X-rays associated with very hot ($\sim 10^8$~K) plasma \citep{koyama1996}. Such powerful explosions in hard X-rays are called protostellar flares. 

Protostellar flares release huge energy ($\sim 10^{34-37}$~erg in the X-ray radiation. The actual released magnetic energy should be much larger) within a time scale of $10^{4-5}$~s \citep{tsuboi2000,montmerle2000,imanishi2003}. 
Statistical studies show that the X-ray detection rate of Class-0/I objects are generally high \citep[$40-60$~\%][]{imanishi2001, getman2007, gudel2007, prisinzano2008, pillitteri2010}, indicating that explosive events ubiquitously occur even just after the birth of protostars. Observations of more evolved (Class-II/III) objects reported that class-III objects produce larger flares than class-II objects, which suggests that the accretion interrupts the energy build-up for flares \citep{getman2008b}. However, actively accreting class-I objects are as bright as class-III objects in the X-ray \citep{pillitteri2010}. This fact motivates us to consider a different mechanism in protostars.

Protostellar flares have been considered as a scale-up version of solar flares. A solar flare is an explosive phenomenon in which the accumulated magnetic energy is violently released by magnetic reconnection \citep{priest2002,benz2008,shibata2016,takasao2016a}. Since solar flares release a fraction of the magnetic energy stored above sunspots with a field strength of a few kG, the flare energy can be estimated as
\begin{align}
E_{\rm flare} &\approx f \frac{B_{\rm spot}^2 L_{\rm spot}^3}{8\pi} \nonumber \\
& \approx 10^{32}~{\rm erg}\left ( \frac{f}{0.1}\right) \left( \frac{B_{\rm spot}}{1000~{\rm G}}\right)^2 \left( \frac{L_{\rm spot}}{0.04 R_\odot}\right)^3,
\end{align}
where $B_{\rm spot}$ is the field strength of sunspots and $f$ is the fraction of magnetic energy that can be released as flare energy \citep{shibata2013}. $E_{\rm flare}\approx 10^{32}$~erg is the energy of the largest solar flare ever observed, and the size $L_{\rm spot}$, defined as the square-root of the total area of sunspots, is consistent with observations by \citet{sammis2000}. Since solar flares occur in the corona, it is meaningful to write the energy with the coronal quantities. Considering that the coronal field strength is smaller than the photospheric value due to expansion, we get
\begin{align}
E_{\rm flare} \approx 10^{32}~{\rm erg}\left( \frac{f}{0.1}\right) \left( \frac{B}{300~{\rm G}}\right)^2 \left( \frac{L_{\rm flare}}{0.1 R_\odot}\right)^3.
\end{align}
Although the result depends on the coronal field strength, this relation suggests that the size of the largest flare is roughly 10\% of the solar radius and this is supported by observations.

We note that X-ray observations can only measure a small fraction of the total released energy \citep[e.g.][]{benz2008,emslie2012}. For instance, \citet{emslie2012} suggest that only $\sim 1~$\% of the total released energy is radiated in X-rays. Although there are uncertainties, the fact from solar observations is that the total released energy should be one or more orders of magnitude larger than the energy in X-rays. Therefore, the total released energy of protostellar flares should be larger than $10^{34-37}$~ergs.

We repeat the above estimation for protostellar flares. The huge flare energy suggests that either strong magnetic fields and/or a large spatial scale are required. The maximum field strength should be at most a few kG even for protostars, otherwise the magnetic field cannot be confined by the gas pressure around the stellar surface \citep[for direct measurements of kG fields for classical T-Tauri stars, see e.g.][]{donati2009}. 
Therefore, the field strength at the footpoints will be comparable to the sunspots, and protostellar flares should have a very large spatial scale to release the observed flare energy:
\begin{align}
E_{\rm flare} \approx 10^{37}~{\rm erg} \left( \frac{f}{0.1}\right)\left( \frac{B}{300~{\rm G}}\right)^2 \left( \frac{L_{\rm flare}}{5 R_\odot}\right)^3.
\end{align}
Since the typical radius of protostars is a few solar radii \citep{baraffe2009,hosokawa2011}, this estimate indicates that the size of protostellar flares is comparable to or larger than the stellar radius. Mechanisms for the energy build-up at such very large scales remain unresolved.


The energy build-up in the case of solar flares is as follows.
Magnetic fields first emerge from the solar interior to form sunspots. Then, sunspots show shearing and/or rotating motions, shearing up coronal magnetic loops  \citep{takasao2015b}. 
This energy build-up may not be expected in the case of protostellar flares, because the spatial scale is larger than the stellar radius and therefore both of the footpoints of magnetic loops may not be anchored by the stellar surface.

What makes the difference between the Sun and a protostar? One crucial factor is the existence of accretion disks.
The large spatial scale suggests that protostellar flares are associated with the dynamical interaction between the protostar and a circumstellar disk. The dynamical evolution of systems in which a protostar has a strong dipolar magnetosphere threading the disk has been investigated \citep{ghosh1979a, koenigl1991,camenzind1990,shu1994,matt2005,ferreira2006}. In this configuration, magnetic loops of the magnetosphere are wound up by the rotating disk (i.e. rotational or gravitational energy of the disk is efficiently converted into magnetic energy). As the twist accumulates, magnetic loops expand to form an electric current sheet inside the loops \citep{lynden-bell1994,lovelace1995}. Magnetic reconnection eventually takes place in the current sheet, and rapidly releases the magnetic energy stored at a scale larger than the stellar radius. Although one end of the foot-points is anchored in the disk, this energy build-up by shear is similar to the case of solar flares. \citet{hayashi1996} investigated the magnetospheric model in detail using a two-dimensional magnetohydrodynamic (2D MHD) simulation \citep[see also][]{goodson1997,hirose1997,uzdensky2004,zanni2013}, and showed that magnetic reconnection naturally accounts for the formation of hot plasma. The accretion process of the star with a magnetosphere has also been studied \citep{miller1997,romanova2009a}. The condition for the magnetospheric accretion has been widely discussed \citep[e.g.][and references therein]{bessolaz2008}.

Although the magnetospheric model is promising, it is possible that actual protostars have no magnetospheres because vigorous disk accretion and strong disk magnetic fields in this phase can destroy the magnetosphere. For instance, if the inner disk has a poloidal magnetic field with the strength of 40~G, the total disk magnetic flux within $\sim 5R_{*}$ is enough to open up the closed stellar dipole fields with the strength of 1~kG through magnetic reconnection. Although there is a large uncertainty in the above estimation, an observation indicates the existence of kG fields there in a highly accreting young star \citep{donati2005}. A recent simulation by \citet{machida2019} shows the existence of strong ($>100$~G) poloidal magnetic fields around a protostar, although the simulation only investigates the very early phase of the star formation (until two thousand years after the birth of the protostar). The strong disk fields can be rapidly transported to the protostar by accretion near/above the disk surface \citep{beckwith2009, takasao2018}, which will also enhance the destruction of the stellar magnetosphere via reconnection. If protostars have no magnetospheres that interact with rotating disks, the energy build-up process in which field lines of the magnetospheres are twisted by the rotating disks will not operate. Does this mean that protostellar flares will not occur without magnetospheres? This is the central question of this paper. We investigated whether protostellar flares can occur in the absence of magnetosphere using a 3D MHD simulation. We report here that protostellar flares can occur even in such a case. This study discusses the relationship between the accretion process and the protostellar flares.

\section{Numerical Setup}
Our numerical setting (including method and initial and boundary conditions) is similar to that in our previous paper \citep{takasao2018}. The differences are 1. a resistivity is included to model reconnection, 2. the disk is magnetized more strongly, 3. the dual energy formalism (the time evolution of the internal energy is also solved) is used to avoid negative pressure in the low plasma $\beta$ regions \citep[e.g.][]{takasao2015b}, and 4. the inner boundary condition is modified. Here we briefly describe our model. For more detailed information, the reader is referred to our previous paper. We solve the 3D resistive MHD equations in a conservative form in spherical coordinates $(r,\theta,\phi)$ using Athena$++$ (J. Stone et al. in preparation). We include a simplified radiative cooling term for the disk material in the energy equation to sustain the initial disk temperature profile. 

We include a resistivity to capture magnetic reconnection. Our resistivity $\eta_{\rm anom}$ is an anomalous resistivity which operates only in the regions with a large relative electron-ion drift velocity $v_{\rm drift}$ (proportional to the current density divided by the mass density) in order to spatially localize the resistivity and realize a fast reconnection \citep{ugai1992}. The functional form is $\eta_{\rm anom} = \eta_0 (v_{\rm drift}/v_{\rm cri}-1)$ in the region where $v_{\rm drift}>v_{\rm cri}$ and the density is 1,000 times smaller than that of the initial inner disk, otherwise $\eta_{\rm anom} = 0$. $v_{\rm cri}$ is the threshold for the drift velocity. We chose the constants $\eta_0$ and $v_{\rm cri}$ so that the current sheets are resolved with at least several meshes. Therefore, the resistivity does not operate in the disk.

The protostar is surrounded by a cold disk threaded by an hourglass-shaped poloidal magnetic field in the initial state. The protostellar mass and radius are $0.5M_\odot$ and $2R_\odot$, respectively. The Keplerian orbital period at the stellar radius is $\sim 0.46$~days. The ratio of the disk pressure scale height to the radius is about 0.14, as in our previous model. The initial plasma $\beta$, the ratio of the gas pressure to the magnetic pressure, on the equatorial plane is constant with radius and set to $10^2$. We expect that the inner part of the accretion disk is more strongly magnetized in the protostellar phase because a fossil magnetic field may still remain there \citep[e.g.][]{machida2019}. The initial disk field strength is approximately 100~G in the innermost region in this study. The initial midplane disk density is $10^{16}~$cm$^{-3}$ at the inner edge, and it decreases to approximately $10^{14}~$cm$^{-3}$ in 23~days after the simulation starts.

We adopt an outgoing boundary condition for the outer boundary. Our inner boundary is a rotating stellar surface which gradually absorbs the accreting mass. The inner boundary is controlled by the boundary condition plus a damping layer (a thin spherical shell around the actual inner boundary). The difference from our previous setting is that the values of the latitudinal and azimuthal components of magnetic fields in the ghost cells are set to zero to improve the numerical stability. The latitudinal component of the velocity is also set to zero in the ghost cells as in our previous paper. The protostar is rotating and the corotation radius is set to $3R_*$, where $R_*$ is the stellar radius, and the magnetic fields threading the stellar surface are also rotated at this velocity. We investigated the influence of the boundary condition by performing a set of 2D simulations with different boundary conditions (zero-value/reflecting boundaries for $v_\theta$ and free/radial-component-only boundaries for magnetic fields), and confirmed that protostellar flares occur in all the cases.

We adopt static mesh refinement to capture turbulence and magnetic reconnection around the disk. The simulation domain is $(0.9R_*, 0, 0)\le (r,\theta,\phi) < (30R_*,\pi,2\pi)$. The radial grid size is proportional to the radius to keep the ratio between the radial and longitudinal grid sizes. We resolve the simulation domain with $120 \times 120 \times 80$ grid cells at the root level (level 0). The refinement level increases toward the midplane of the inner disk. The maximum level is 2 ($r<6R_*$).

\section{Numerical Results}
\begin{figure}
\epsscale{1.3}
\plotone{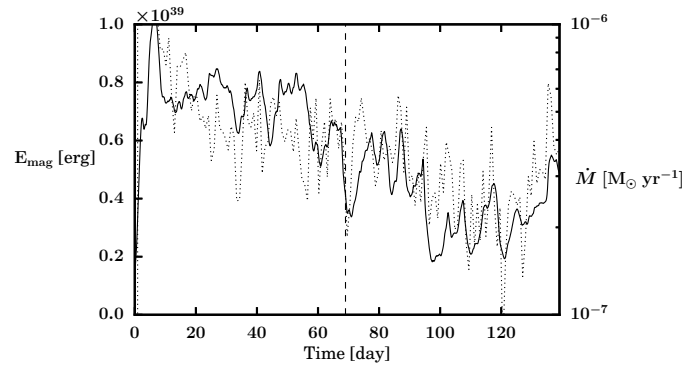}
\caption{Temporal evolution of the total magnetic energy integrated within the radius of $5R_*$ (solid) and the rate of mass accretion onto the protostar (dotted). The vertical dashed line indicates the timing of the flare shown in Figures~\ref{fig:flare3d} and \ref{fig:flare-midplane}.}\label{fig:emag}
\end{figure}

Figure~\ref{fig:emag} shows the evolution of the magnetic energy around the protostar and the accretion rate. The time-averaged accretion rate is $\sim 3\times 10^{-7}~M_\odot~{\rm yr^{-1}}$. Within the distance of five stellar radii, the protostar stores magnetic energy of $\sim 10^{39}$~erg, which is sufficient to produce large protostellar flares. The plot indeed displays many events of sudden release of magnetic energy. 
The initial sharp increase in the magnetic energy ($t<10$~days) is due to quick accumulation of magnetic flux from the disk.
Such efficient transport of magnetic flux is also found in our previous simulation \citep{takasao2018} and caused by a rapid accretion near and above the disk surface where the plasma $\beta$ is lower and the magnetic torque can transport angular momentum more efficiently than around the equatorial plane \citep{matsumoto1996,beckwith2009,zhu2018}. As discussed later, the energy accumulation after energy release events is also mediated in this manner. We observe fast funnel-accretion found in our previous paper \citep{takasao2018}.

\begin{figure*}
\epsscale{1}
\plotone{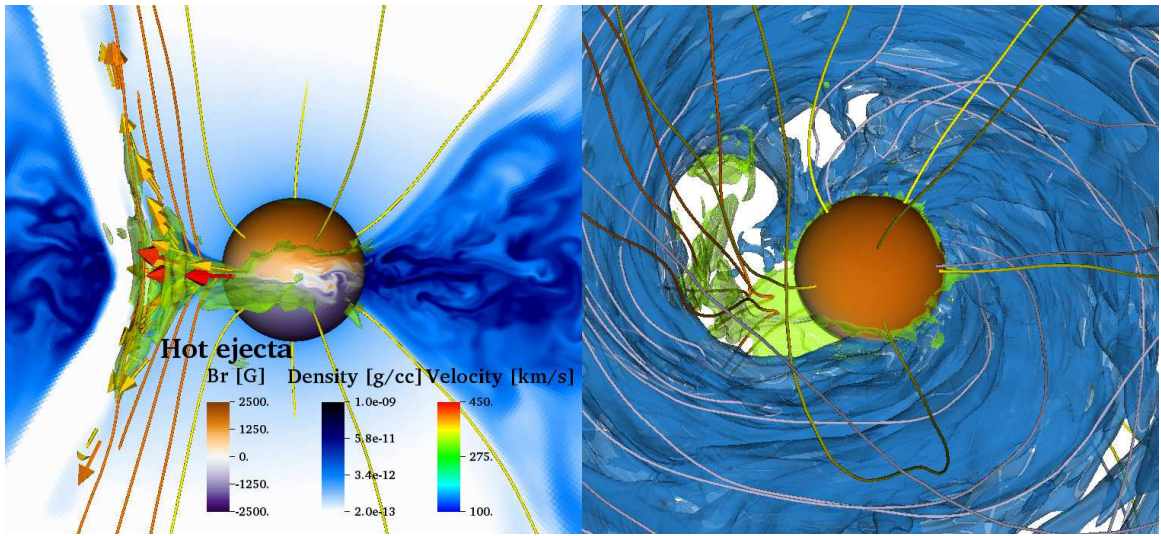}
\caption{3D images of a protostellar flare which occurred at $t\sim68$~days. The protostellar surface is colored by the value of the radial component of magnetic fields. In the left panel (side view), the density is shown by the poloidal slice. The blue isosurface in the right panel (top view) indicates the density of $3\times10^{-11}$~g~cm$^{-3}$. A hot plasma ejection is indicated by the isosurface of the temperature (yellow, the temperature is $\sim 5\times 10^{6}$~K). Reconnected magnetic field lines are colored by orange, while the stellar and disk field lines are by yellow and purple, respectively. Arrows denote velocity vectors.}\label{fig:flare3d}
\end{figure*}

Figure~\ref{fig:flare3d} displays an example of the explosive events which occurred at $t\sim 68$~days. We find a hot ($\sim10^{7}$~K) plasma ejection from the protostar with a velocity of $\sim$450~km~s$^{-1}$ (yellow-colored region). A portion of the reconnection outflow is refracted along the reconnected fields to form a hot bipolar outflow. We observe many explosions like this event and interpret them as protostellar flares. The averaged energy conversion efficiency is approximately $L_{\rm flare}\sim 0.1 \dot{E}_{\rm grav}$, where $L_{\rm flare}$ and $\dot{E}_{\rm grav}=-GM_* \dot{M}/R_*$ are the time-averaged flare luminosity and the energy release rate by accretion, respectively.


\begin{figure*}
\epsscale{1.0}
\plotone{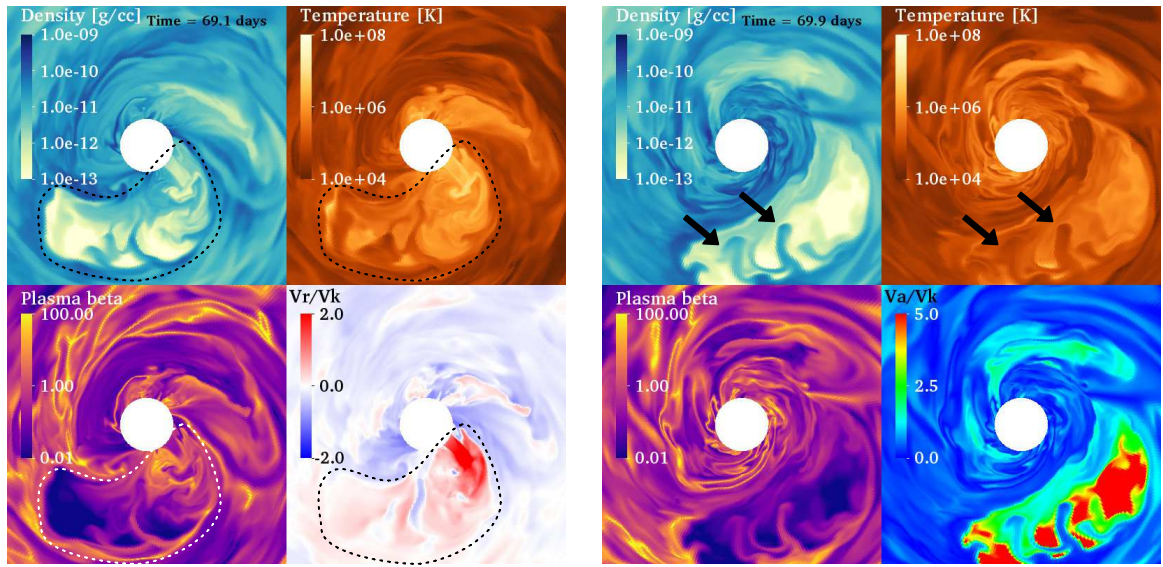}
\caption{The distributions of physical quantities at the midplane at two different times. Left: the density (top left), the temperature (top right), the plasma $\beta$ (bottom left), and the radial velocity normalized by the Keplerian velocity at the stellar surface (bottom right) at time $69.1$~days. The hot plasma ejection (indicated by the dashed curves) is driven by the protostellar flare shown in Figure~\ref{fig:flare3d}. Right: same as the left panels but at a later time of $69.9$~days. The bottom right panel shows the ratio of the Alfv\'en speed to the local Keplerian velocity. The arrows in the density and temperature maps indicate finger-like structures caused by the magnetic Rayleigh-Taylor instability. An animation of this figure is available. The sequence starts at time 65.8~days and ends at time 75.1~days (20 Keplerian orbital periods at the stellar surface).}\label{fig:flare-midplane}
\end{figure*}

Protostellar flares are driven by magnetic reconnection in a way similar to solar flares. The hot ejecta seen in Figure~\ref{fig:flare3d} is a manifestation of magnetic reconnection.
Protostellar flares are so powerful that they significantly affect the disk. Figure~\ref{fig:flare-midplane} displays the disk structure at the equatorial plane at two successive times. The protostellar flare in Figure~\ref{fig:flare3d} appears as a tenuous hot void in this figure (see the region indicated by the dashed curves in the left panel). The void corresponds to a cross-section of the bundle of reconnected magnetic fields. The void carries strong magnetic fields away (orange-colored field lines in Figure~\ref{fig:flare3d}), removing large-scale poloidal fields from the protostar. This is why the plasma $\beta$ in the void is much lower than unity. The density is small because the reconnected field lines come from the tenuous protostellar corona. 

The magnetic fields in the void diffuse out as time progresses because of the magnetic interchange instability. The right panel of Figure~\ref{fig:flare-midplane} shows the result 0.8~days after the time in the right panel. We can find that finger-like structures are developing in the void (indicated by arrows), which are manifestation of the instability. The instability can appear only in regions where the Lorentz force is operating significantly against gravity \citep{stehle2001}. In fact, in the ejecta the plasma $\beta$ is much smaller than unity, and the magnetic energy density is larger than the gravitational energy density (indicated by the ratio of the Alfv\'en speed to the local Keplerian velocity, the bottom right panel). We also confirmed that the instability condition is satisfied by checking the gradient of the magnetic field strength divided by the surface density. The accretion is initially driven by magneto-rotational instability \citep[MRI][]{balbus1991}, but MRI is suppressed later due to strong magnetization. The accretion rate around the protostar seems to be sustained by a combination of the magnetic interchange instability and the disk outflow (not shown here).

\begin{figure*}
\epsscale{1.1}
\plotone{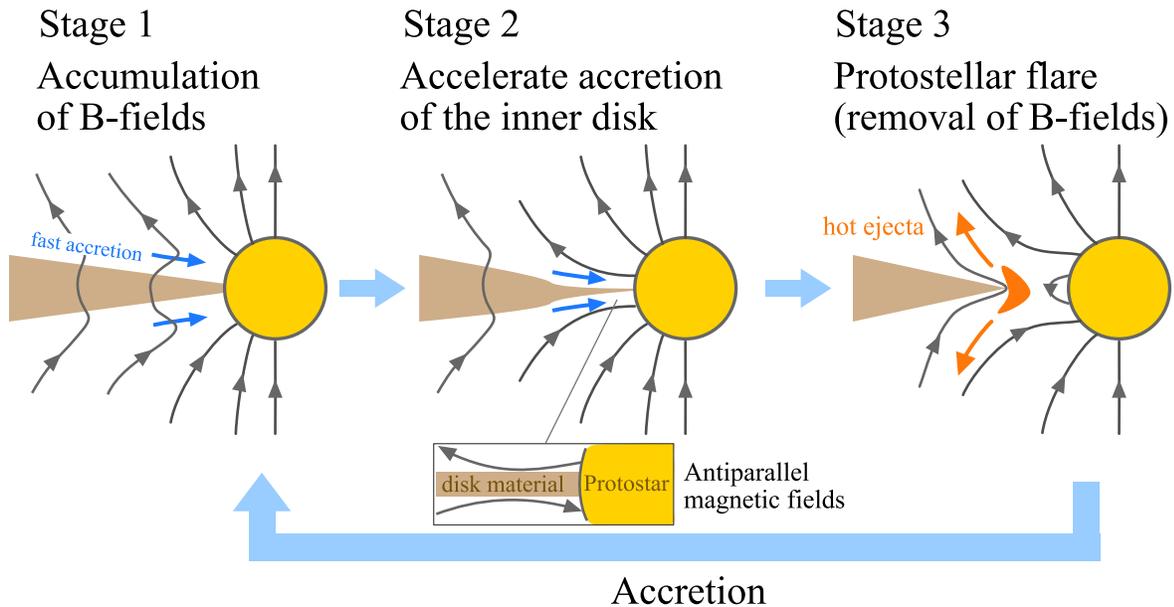}
\caption{Schematic diagram of the newly found mechanism of the protostellar flares.}\label{fig:flare-description}
\end{figure*}

Why do magnetic fields which accumulate around the protostar overflow into the disk? We describe the process with a schematic diagram in Figure~\ref{fig:flare-description}. The protostar acquires large-scale magnetic fields from the disk by accretion (stage~1). When the stellar magnetic fields become strong enough, the stellar fields expand toward the disk. Since the stellar rotation is slower than the Keplerian rotation in this study, the stellar fields remove the angular momentum of the disk gas and accelerate the disk surface accretion near the protostar (stage~2). Once the inner disk is cleared up, magnetic fields of the stellar north and south poles contact to reconnect around the equatorial plane, producing a flare (stage~3). In this way, a fraction of the stellar fields are removed and go into the disk. 

We will describe why protostellar flares repeatedly occur. Although the flare drives an outgoing ejection, the ejection is spatially limited in the azimuthal direction (Figure~\ref{fig:flare-midplane}) and therefore the accretion continues. In addition, the magnetic fields ejected by reconnection are moving radially, and therefore they behave as an obstacle for surrounding rotating disk materials. Disk materials lose their angular momenta when interacting with the magnetic fields. The disk accretion is enhanced as a result (see the correlation between the magnetic energy and the accretion rate in Figure~\ref{fig:emag}). Thus, the inner density gap is filled up and the disk accretion can accumulate magnetic fields around the protostar again.

\begin{figure}
\epsscale{1.}
\plotone{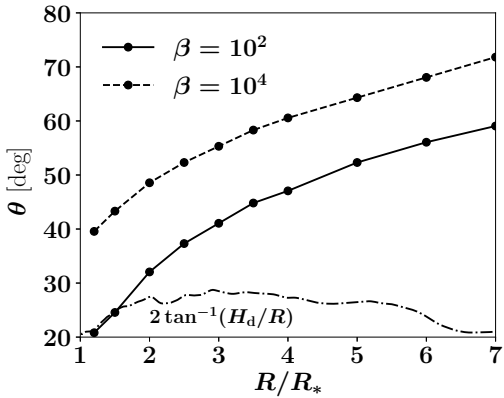}
\caption{Radial distribution of the angle of the pressure balance point. The angle is measured from the equatorial plane. The solid line denotes the result in the simulation of this study. For comparison, we also plot the result from the weaker field case in \citet{takasao2018} ($\beta=10^4$, dashed line). We also plot $2\tan^{-1}{(H_{\rm d}/R)}$ as an indication of the disk opening angle, where $H_{\rm d}$ is the width of the fitting Gaussian profile (dashed-dotted line). The fitting is performed at each radius up to the heights at which the density is 10\% of the midplane gas. Note that the initial disk opening angle is $\sim16^\circ$.}\label{fig:balance-angle}
\end{figure}

We investigated the size of the influence region of the stellar fields. The stellar fields are confined in the polar regions by the sum of the gas and magnetic pressures of the disk \citep[see][]{takasao2018}, which means that the opening angle of the stellar magnetic funnel is determined by the pressure balance in the latitudinal direction. If the balance point reaches close to or inside the disk, the stellar fields have significant impact on the disk accretion. Figure~\ref{fig:balance-angle} shows the angle of the balance point, where the angle is measured from the equatorial plane. We also measured the disk opening angle numerically (see the caption in Figure~\ref{fig:balance-angle}). The figure shows that in this simulation (solid line, the initial plasma $\beta=10^2$) the balance point comes into or close to the disk within the radius of $1.5-2R_{*}$. Therefore, the stellar fields significantly affect the disk within this radius. In a weaker field case, on the other hand, the balance point is located well above the disk surface, and we do not observe strong flares.

\section{Summary and Discussion}
We presented the results of a global 3D MHD simulation of protostellar flares under the assumption that the protostar does not have a magnetosphere initially. We found that protostellar flares repeatedly occur even in the absence of a stellar magnetosphere. The protostar accumulates huge magnetic energy by getting large-scale magnetic fields from the disk. The stored magnetic energy is released by magnetic reconnection to produce protostellar flares. A fraction of the stellar fields are removed as a result of reconnection. 

Protostellar flares release a large amount of magnetic energy ($\sim 10^{38}$~ergs) because of their large spatial scale. When we consider the energy partition of flares, only a small fraction of this energy will be radiated in X-ray. If the energy conversion rate is 1\%, the flare energy observed in the simulation can account for protostellar flares with the energy of $\sim 10^{36}$~ergs estimated from X-ray observations. 

We compare our ``flux removal model" with the magnetospheric model by \citet{hayashi1996}. In both models, the energy build-up and release are controlled by the surrounding disks, and flares can occur repeatedly to produce hot ejecta. However, some big differences can be found. In the magnetospheric model, flares are triggered when field lines of the magnetosphere are wound up by approximately one rotation \citep{lynden-bell1994}, but observations do not support this \citep{imanishi2001}. Reconnection occurs well above the disk, and therefore the disk is less affected by ejecta. In our model, the energy build-up is the magnetic field accumulation by accretion. The time interval between two successive flares (flare interval) is determined by the star-disk interaction and more than 10 orbital periods at the stellar surface. Reconnection happens around the midplane, and therefore largely disturbs the inner disk. Inhomogeneity in the azimuthal direction is also prominent because of the localized reconnection and the magnetic interchange instability in the ejecta. The accretion rate is not totally quenched by flare ejecta due to the inhomogeneity.


Protostellar flares occur with a typical period of approximately 20-30 orbital periods (or 10-15~days) in this study. This time scale is consistent with the idea that protostellar flares occur once the protostellar fields clear up the inner disk. The flare interval will be characterized by the timescale at which the inner disk is cleared up by the interaction between the stellar magnetic fields and the disk surface. From Figure~\ref{fig:balance-angle} the interaction is important within the radius of $\sim2R_*$. We define the accretion rate around the disk surfaces as $\dot{M}_{\rm surf} = 4\pi R \rho_{\rm surf} v_{\rm r}w$, where $\rho_{\rm surf}$ is the density around the disk surface and $w$ is a typical width of the disk surface accretion (here we take $w=H_{\rm p}$). Then the timescale $t_{\rm clear}$ is estimated as $M(R<2R_*)/\dot{M}\approx \pi R^2 \rho_{\rm mid} \cdot 2H_{\rm p} / 4\pi R \rho_{\rm surf} v_{\rm r} w = (1/2)(2\pi R/v_{\rm r})(\rho_{\rm mid}/\rho_{\rm surf})$. Our numerical simulation indicates that $2\pi R/v_{\rm r}\sim 5 t_{\rm K}$, and $\rho_{\rm mid}/\rho_{\rm surf}\sim 10$ at $R=2R_*$, where $t_{\rm K}$ is the Keplerian orbital period at $R_*$ ($\sim 0.46$~days). Therefore, we obtain that $t_{\rm clear}\sim 25 t_{\rm K}\sim 12~$days, which is consistent with the numerical result.

A significant amount of magnetic fields is carried to the protostar by accretion outside the disk body as commonly seen in many simulations \citep[e.g.][]{beckwith2009,suzuki2014,takasao2018}. As a result, the accretion flows around the protostar become essentially the same as so-called magnetically chocked accretion flows (MCAFs) \citep{mckinney2012b}. MCAFs are similar to the ``magnetically arrested disc" (MAD) flows \citep{narayan2003}, widely discussed in the context of black hole accretion and MRI is suppressed due to strong magnetization \citep{white2019}. \citet{mckinney2012b} report that the magnetic flux redistribution is mediated by the instabilities once magnetic flux has accumulated up to a saturation point. Here we emphasize the importance of magnetic reconnection. As shown in Figures~\ref{fig:flare-midplane} and \ref{fig:flare-description}, the magnetic flux removal via reconnection also leads to the redistribution. Protostellar flares will largely affect the disk magnetic field evolution in the innermost region. Reconnection similar to this but around a black hole is also discussed by \citet{komissarov2005}.

Less attention has been paid to the flaring activity in the context of the protostellar evolution. However, our results suggest that the disk material can be significantly heated up by protostellar flares before accreting onto the star, which may affect the stellar evolution through the change in the entropy carried into the protostar \citep[e.g.][]{hosokawa2011,kunitomo2017}. The heating may also contribute to the formation of chondrules which are believed to be the building blocks of planets \citep[see also][]{nakamoto2005}. 
The magnetic flux removal will be important for the transition of the stellar magnetic structure; the stellar magnetic fields should be dominated by fossil, open poloidal fields in the protostellar phase. However, the magnetic fields of evolved stars like CTTSs are dominated by multipolar, dynamo-generated fields. 
Our results indicate that protostellar flares can occur even before the stellar dynamo starts. Therefore, studying protostellar flares may enable us to probe the very initial state after the birth of protostars.
We will explore the impact on the star and planet formation in future.

\acknowledgments
We thank Drs. K. Shibata and S. Inutsuka for fruitful discussion. S.T. acknowledges support by the Research Fellowship of the Japan Society for the Promotion of Science (JSPS). This work was supported in part by JSPS KAKENHI grant No. 16J02063 (S.T.) and 17H01105 (T.K.S.). Numerical computations were carried out on the Cray XC50 at the Center for Computational Astrophysics, National Astronomical Observatory of Japan. Test calculations were carried out on the XC40 at the Yukawa Institute for Theoretical Physics in Kyoto University. This research was also supported by MEXT as ``Exploratory Challenge on Post-K computer" (Elucidation of the Birth of Exoplanets [Second Earth] and the Environmental Variations of Planets in the Solar System).


\bibliography{stardisk_takasao}


\end{document}